\documentstyle[12pt,aasms4]{article}

\def\vA{v_{\scriptscriptstyle A}}
\def\tA{t_{\scriptscriptstyle A}}
\def\tp{t_{\scriptscriptstyle p}}
\def\grad{\nabla_{\scriptscriptstyle\perp}}
\def\lapl{\grad\sp2}
\def\dt{\partial_t}
\def\dx{\partial_x}
\def\dy{\partial_y}
\def\dz{\partial_z}
\def\zz{\hat{\bf z}}
\def\l0{l_0}
\def\Eqn#1{Eqn (\ref{eq:#1})}
\def\Eqs#1#2{Eqs (\ref{eq:#1})-(\ref{eq:#2})}
\begin{document}

\title{Scaling law for the heating of solar coronal loops} 
\author{Pablo Dmitruk\altaffilmark{1} and Daniel O. G\'omez\altaffilmark{2,3}}
\affil{Departamento de F\'\i sica, Facultad de Ciencias Exactas y 
Naturales, \\ Universidad de Buenos Aires,\\ Ciudad Universitaria, 
Pabell\'on I, 1428 Buenos Aires, Argentina}

\altaffiltext{1}{Fellow of Universidad de Buenos Aires}
\altaffiltext{2}{Also at the Instituto de Astronom\'\i a y F\'\i sica
del Espacio, Buenos Aires, Argentina}
\altaffiltext{3}{Member of the Carrera del Investigador, CONICET, Argentina}

\authoremail{dmitruk@df.uba.ar}

\begin{abstract}
We report preliminary results from a series of numerical simulations 
 of the reduced magnetohydrodynamic equations, used to describe the 
dynamics of magnetic loops in active regions of the solar corona. 
A stationary velocity field is applied at the photospheric boundaries 
to imitate the driving action of granule motions. 

A turbulent stationary regime is reached, characterized by a 
broadband power spectrum $E_k\simeq k^{-3/2}$ and heating rate levels 
compatible with the energy requirements of active region loops.
A dimensional analysis of the equations indicates that their solutions 
are determined by two dimensionless parameters: the Reynolds number 
and the ratio between the Alfven time and the photospheric turnover 
time. From a series of simulations for different values of this 
ratio, we determine how the heating rate scales with the physical parameters 
of the problem, which might be useful for an observational test 
of this model.

\end{abstract}

\section{Introduction}
 Solar coronal loops are likely to be heated by ohmic dissipation of 
field-aligned electric currents. These currents are driven by 
photospheric motions which twist and shear the magnetic fieldlines 
at the loop footpoints. The generation of small scales in the spatial 
distribution of these currents has been observed and studied in recent 
simulations (\cite{Mik89}, \cite{Lon94}). Small scale currents are required 
by most coronal heating theories, since they produce a major enhancement of 
the energy dissipation rate. The development of magnetohydrodynamic (MHD) 
turbulence in coronal loops has recently been proposed (\cite{Gom92}, 
\cite{Hey92}, \cite{Ein96}, \cite{Hen96}, \cite{Dmi97} ) as a likely 
mechanism to provide such enhancement, since the energy being pumped by 
footpoint motions is naturally transferred to small scale structures by 
the associated energy cascade.

In the present paper we numerically test this scenario, describing 
the dynamics of a coronal loop through the reduced MHD (RMHD) approximation.
In \S 2 we briefly describe the RMHD approximation and perform a 
dimensional analysis of the equations, which leads to an interesting 
prediction for the functional dependence of the heating rate with the 
physical parameters 
of the problem. The details of the numerical simulations are summarized 
in \S 3 and an overview of the results is given in \S 4. A quantitative 
scaling law for the heating rate is derived in \S 5, and the relevant 
results of this paper are listed in \S 6.

\section{Dimensional analysis of the RMHD equations}

When a coronal loop with an initially uniform magnetic 
field ${\bf B} = B_0 \zz$ is driven by photospheric motions at 
its footpoints, a rather complex dynamical evolution sets in, which 
is described by velocity and magnetic fields. For incompressible and 
elongated loops, i.e. loops of length $L$ and transverse section 
$(2\pi\l0)\times (2\pi\l0)$ such that $\l0\ll L$, the velocity (${\bf v} = 
{\bf v} (x,y,z,t)$) and magnetic ${\bf B} = B_0\zz\ +\ {\bf b} (x,y,z,t)$) 
fields can be written in terms of a stream function $\psi = \psi (x,y,z,t)$ 
(i.e. ${\bf v} = \grad\times (\psi\zz)$) and a vector potential $a = a 
(x,y,z,t)$ (i.e. ${\bf b} = \grad\times (a\zz)$), respectively. The 
evolution of the fields $a$ and $\psi$ is determined by the RMHD 
equations (\cite{Str76}):
\begin{equation}
   \dt a  = \vA \dz \psi + [ \psi , a ] + \eta \lapl a
\label{eq:dta}
\end{equation}
\begin{equation}
   \dt w = \vA \dz j + [ \psi , w ] - [ a , j ] + \nu \lapl w
\label{eq:dtw}
\end{equation}
\noindent
where $\vA = B_0/\sqrt{4\pi\rho}$ is the Alfv\'en speed,
$\nu$ is the kinematic viscosity and $\eta $ is the plasma resistivity.  
The quantities $ w=-\lapl\psi $ and  $j=-\lapl a \ $ are the $z$-components 
of vorticity and electric current density, respectively. The non-linear terms 
are standard Poisson brackets, i.e. $[u,v]=\dx u\dy v - \dy u\dx v$. We 
assume periodicity for the lateral boundary conditions, and specify the 
velocity fields at the photospheric boundaries,
\begin{equation}
  \psi (z = 0) = 0,\ \ \ \ \ \ \psi (z = L) = \Psi (x,y)
\label{eq:psiphot}
\end{equation}
where $\Psi (x,y)$ is the stream function which describes stationary and 
incompressible footpoint motions. The strength of 
this external velocity field is therefore proportional to a typical 
photospheric velocity $V_p$.

To transform \Eqs{dta}{dtw} into their dimensionless form,
we choose $\l0$ and $L$ as the units for transverse and longitudinal distances.
Since the dimensions of all physical quantities involved in these equations 
can be expressed as combinations of {\it length} and {\it time}, let us 
choose  $\tA\equiv L/\vA$ as the time unit.
The dimensionless RMHD equations are:
\begin{equation}
   \dt a  = \dz \psi + [ \psi , a ] + {1 \over S} \lapl a
\label{eq:dta2}
\end{equation}
\begin{equation}
   \dt w = \dz j + [ \psi , w ] - [ a , j ] + {1 \over R} \lapl w
\label{eq:dtw2}
\end{equation}
\noindent
where $S= {\l0^2\over{\eta\tA}}$ and $R= {\l0^2\over{\nu\tA}}$ are 
respectively the magnetic and kinetic Reynolds numbers. Hereafter, we 
will consider the case $S = R$, and thus the (common) Reynolds number will be 
the only dimensionless parameter explicitly present in \Eqs{dta2}{dtw2} . 
However, the boundary condition (\Eqn{psiphot}) brings an extra 
dimensionless parameter into play, which is the photospheric velocity $V_p$ 
divided by our velocity units, i.e. $\l0 /\tA$. Since $V_p = \l0 /\tp$ 
($\tp$: photospheric turnover time), this velocity ratio is equivalent to
the ratio between the Alfven 
time $\tA$ and the photospheric turnover time $\tp$.  Therefore, from 
purely dimensional considerations, we can derive the following important 
result: {\it for any physical quantity, its dimensionless version} 
$Q$ {\it should be an arbitrary function of the only two dimensionless 
parameters of the problem, i.e.}
\begin{equation}
  Q = {\cal F} ( Q_1 , Q_2 ),\ \ \ \ Q_1 = {\tA\over\tp},\ \ \ Q_2 = S
\label{eq:Q}
\end{equation}

For instance, let us consider the important case of the heating rate 
per unit mass, i.e. $\epsilon /\rho$. Its dimensionless version is,
\begin{equation}
  Q = {\epsilon\over\rho} {\tA^3\over\l0^2} = {\cal F} ( {\tA\over\tp} , S )
\label{eq:Qeps1}
\end{equation}

One of Kolmogorov's hypothesis in his theory for stationary turbulent 
regimes at very large Reynolds numbers (\cite{Kol41}), states that
 the dissipation rate is independent of the Reynolds number (see also 
\cite{Fri96}). In the next section we show that externally driven coronal 
loops eventually reach a stationary turbulent regime. Therefore, hereafter 
we assume that the dependence of the dissipation rate $\epsilon$ with 
the Reynolds number $S$ in \Eqn{Qeps1} can be neglected. For moderated
values of the Reynolds numbers as the ones considered here, we observed
only a mild dependence of the dissipation rate with S (see also 
\cite{Hetal96}), in accordance with Kolmogorov's assumption. Therefore, 

\begin{equation}
   \epsilon = {\rho\l0^2\over\tA^3} {\cal F} ( {\tA\over\tp} )
\label{eq:Qeps2}
\end{equation}
 
We performed a sequence of numerical simulations for different 
values of the ratio ${\tA\over\tp}$ to determine the function ${\cal F}$.

\section{Numerical simulations}

For the numerical simulations of \Eqs{dta2}{dtw2} , $\psi$ and $a$ are 
expanded in Fourier modes in each $(x,y)$ plane ($0 \le x,y \le 2\pi$ 
and $0\le z \le 1$). The equations for the coefficients $\psi_{\bf k} 
(z,t)$ and $a_{\bf k} (z,t)$ are time-evolved using a semi-implicit scheme. 
Linear terms are treated in a fully implicit fashion, while nonlinear terms 
are evolved using a predictor-corrector scheme. Also, nonlinear terms are 
evaluated following a $2/3$ fully dealiased (see \cite{Can88}) 
pseudo-spectral technique (see also \cite{Dmi98}). To compute $z$-derivatives 
we use a standard method of finite differences in a staggered regular grid 
(see for instance \cite{Str76}). 

We model the photospheric boundary motion in \Eqn{psiphot} as
$ \Psi_{\bf k} = \Psi_0 =$ constant inside the ring $3 < k\l0 < 4$, and 
$ \Psi_{\bf k} = 0$ elsewhere. This choice is intended to simulate a 
stationary and isotropic pattern of 
photospheric granular motions of diameters between $2\pi\l0/4$ and 
$2\pi\l0/3$. We chose a narrowband and non-random forcing to make sure 
that the broadband energy spectra and the signatures of intermittency that 
we obtained (see below) are exclusively determined by the nonlinear nature 
of the MHD equations. The normalization factor $\Psi_0$ is 
proportional to $\tA / \tp$, as mentioned above.

\section{Development of MHD turbulence}

We performed a sequence of numerical simulations of \Eqs{dta2}{dtw2} with 
$192\times 192\times 32$ gridpoints, $S = 1500$ and different values 
of the ratio $\tA /\tp$ in the range $[ 0.01 , 0.15 ]$. The typical 
behavior of the heating rate as a function of time, is shown in Figure 1 
for the particular case of $\tA /\tp = 0.064$. After an initial 
transient, the heating rate is seen to approach a stationary level.
 This level is fully consistent with the heating requirements for
coronal active regions of  $10^7\ erg\ cm^{-2}\ seg^{-1}$ (\cite{Wit77}).
The intermittent behavior of this time series, which is typical of 
turbulent systems, is also ubiquitous in all our simulations. Once 
the stationary regime is reached in each of the simulations, we determine 
the mean value and rms of the departure of the series from the mean.
%
%

A stronger indication of the presence of a turbulent regime, is the 
development of a broadband energy power spectrum, which behaves like 
$E_k\simeq k^{-3/2}$ for both two 
and three dimensional MHD turbulence (Kraichnan spectrum). 
To compute the energy power spectrum, we performed a single simulation 
with better spatial resolution ($384\times 384\times 32$), to allow for 
a more extended inertial range. The parameter for this simulation 
is $\tA /\tp = 0.064$ and $S = 5000$. The spectrum shown in Fig. 2 
was taken at $t = 5\tp $, i.e. well in the stationary regime. This higher   
spatial resolution is still insufficient for the formation of a 
broad inertial range to determine the slope of these spectra with 
confidence. However, the observed slope is consistent with a Kraichnan 
spectrum, i.e. $E_k\simeq k^{-3/2}$. Also, note that the kinetic and 
magnetic power spectra reach equipartition at large wavenumbers, even 
though the total kinetic energy is much smaller than the total magnetic 
energy. 
%
%

Another important consequence of the energy power spectrum displayed 
in Fig. 2, is the fact that smaller scales dissipate more energy 
than the larger scales, since $\epsilon_k\propto\ k^2 E_k\simeq k^{1/2}$. 
Three dimensional simulations of loops driven by random footpoint motions 
have shown that energy dissipation preferentially occurs in current sheets 
which form exponentially fast (\cite{Mik89}, \cite{Lon94}). Within the 
framework of MHD turbulence, two dimensional simulations also show the 
ubiquitous presence of current sheets evolving in a rather dynamic fashion 
(\cite{ML86}, \cite{BW89}, \cite{Dmi98}). The presence of elongated current 
sheets as the most common dissipative structures is also apparent in these 
simulations (also \cite{Hen96}). Figure 3 shows the spatial distribution of 
the electric current density $j (x,y,z,t)$ for our $384\times 384\times 32$ 
simulation ($\tA /\tp = 0.064$ and $S = 5000$). Intense currents going 
upward (downward) are shown in white (black) at various planes $z = 
constant$, and for $t = 5\tp $. Although the reconnection in these turbulent 
regimes is expected to be fast (\cite{Pri92}, \cite{Hen96}, \cite{Mil99}), 
the Reynolds numbers attained by current simulations cannot provide a definite 
answer to this question.

%
%

\section{Determination of the heating rate scaling exponent}
One of the main goals of this paper is to derive the functional dependence 
of the heating rate with the ratio between the two relevant timescales (see 
\Eqn{Qeps2}). To this end, we performed various simulations for 
different values of the parameter $\tA /\tp$.
Figure 4 shows that this functional dependence can be adequately fit by 
a power law, i.e.

\begin{equation}
   \epsilon = {\rho\l0^2\over\tA^3} ( {\tA\over\tp} )^s, \ \ \ \ \ \ \ \ \ \ \ \ \ \ \ s = 1.51 \pm 0.04
\label{eq:Qeps3}
\end{equation}

 Each diamond corresponds to the mean value of the heating rate for a 
particular simulation in its stationary regime. The error bars correspond 
to the rms value of the departure from the mean. The slope in this plot 
corresponds to the parameter $s$ quoted in \Eqn{Qeps3}, which was derived 
following a minimum squares procedure.
%
%

This result is fully consistent with the prediction arising from a two 
dimensional MHD model (\cite{Dmi97}), $s_{\scriptscriptstyle 
2D} = {3\over2}$, which is assumed to simulate the dynamics of a generic 
transverse slice of a loop. This coincidence between both scalings suggests 
that two dimensional simulations provide an adequate description of the 
dynamics  of coronal loops (see also \cite{Hen96}), provided that the 
external forcing in \Eqs{dta}{dtw} is given by $\vA \dz \psi \approx 
{{\l0 V_p}\over\tA}$ and $\vA \dz j \approx 0$ (\cite{Ein96}, \cite{Dmi97}).
\section{Discussion and conclusions}
In the present paper we report preliminary results from a series of numerical 
simulations of the RMHD equations driven at their footpoints by stationary 
granule-size motions. The important results are summarized as follows:
\begin{itemize}
\item[(1)] After a few photospheric turnover times, the system relaxes to 
 a stationary turbulent regime, with dissipation rates consistent with 
the heating requirements of coronal active regions ($10^6\ -\ 10^7\ erg\ 
cm^{-2}\ s^{-1}$).

\item[(2)] A dimensional analysis of the RMHD equations shows that the 
 dimensionless heating rate does only depend on the ratio $\tA /\tp$ and 
 the Reynolds number. However, for turbulent systems at very large 
 Reynolds numbers, the heating rate is likely to be independent of the 
 Reynolds number. From a series of numerical simulations, we find that 
 the heating rate scales with $\tA /\tp$ as $\epsilon\simeq {{\rho\l0^2} 
 \over\tA^3}({\tA\over\tp})^{3/2}$, i.e.
\begin{equation}
 \epsilon\simeq {{\rho^{1\over4}(B_0 V_p )^{3\over2}}\over L} ({\l0\over L})^{1\over2}
\label{eq:fin}
\end{equation}
 which might be useful for an observational test of this theoretical 
 framework of coronal heating.

\item[(3)] The dissipative structures observed in our simulations are 
 current sheets elongated along the axis of the loop. The spatial and 
 temporal distribution of these structures is rather intermittent, as 
 expected for turbulent regimes. We associate these dissipation events 
 taking place inside coronal loops to the {\it nanoflares} envisioned 
 by \cite{Par88} , as a likely scenario for coronal heating.
\end{itemize}



\newpage
\figcaption[fig_1.ps]{Dimensionless heating rate as a function of time 
(in units of $\tA$) for $\tA /\tp = 0.064$ and $S=1500$. The thin trace 
corresponds to viscous heating.
\label{f1}}

\figcaption[fig_2.ps]{Energy power spectrum for a $384\times 384\times 32$ simulation 
 at $t = 5\tp$. The thin trace corresponds to kinetic energy. A straight line 
corresponding to a slope $-3/2$ is displayed for reference.
\label{f2}}

\figcaption[fig_3.ps]{Halftones of the function $j (x, y)$ at the transverse planes 
$z = 0, 0.33, 0.66, 1$ and for $t=5 \tp$. Upflowing currents are white and 
downflowing currents are black.
\label{f3}}

\figcaption[fig_4.ps]{Dimensionless heating rate as a function of 
$\tA /\tp$. Each diamond corresponds to a different simulation and the 
full line corresponds to the slope determined from a minimum squares 
fit.
\label{f4}}

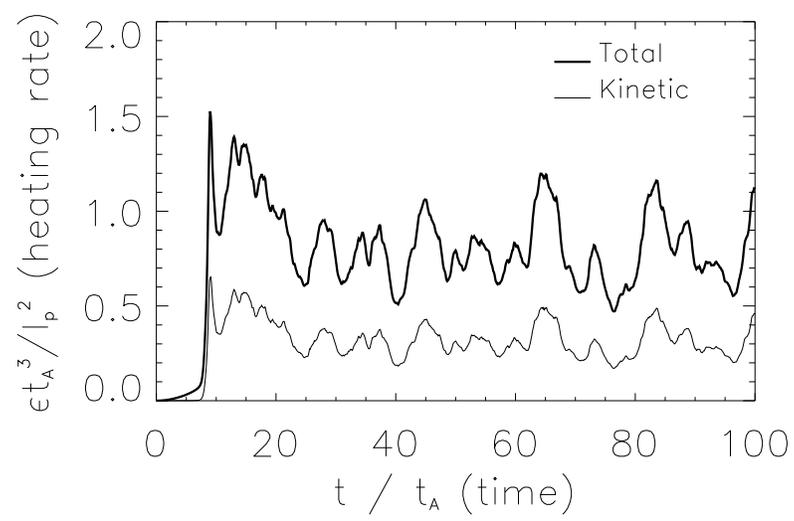
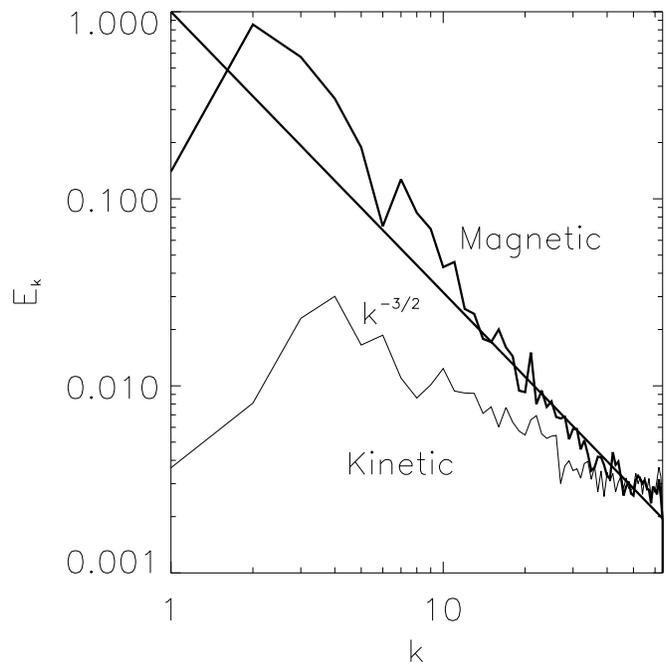
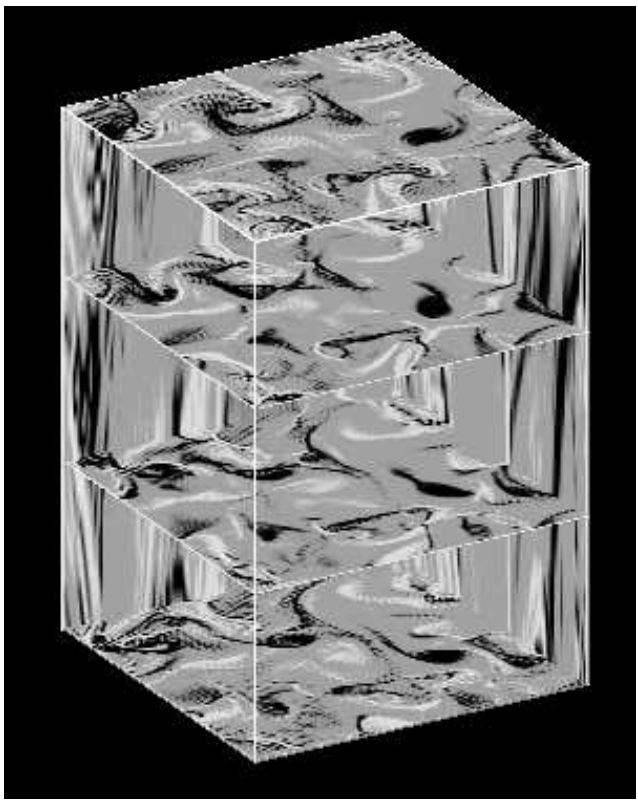
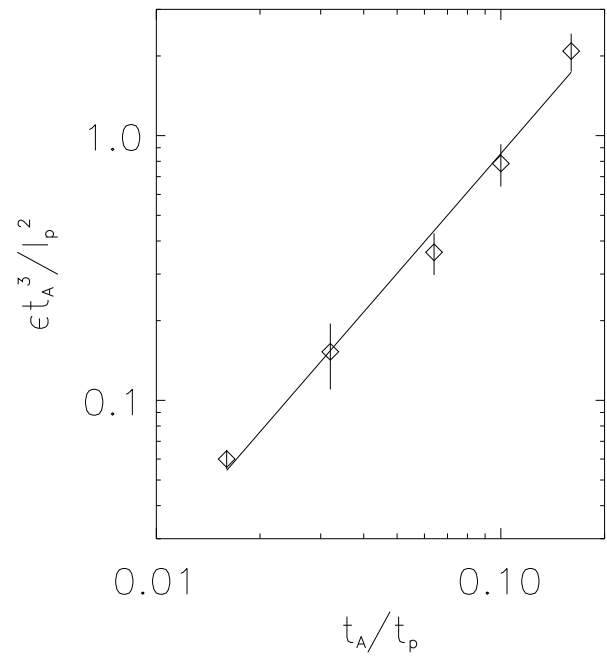
\end{document}